\documentclass[twocolumn]{aastex631}
\shortauthors{Wu et al.}
\shorttitle{Diffuse Neutrinos from MADs}

\usepackage{graphicx}
\usepackage{amsmath,amssymb}
\usepackage[T1]{fontenc}
\begin{document}

\title{PeV Diffuse Neutrinos from Magnetically Arrested Disks in Low-Luminosity Active Galactic Nuclei}

\author[0000-0002-6292-057X]{Fan Wu}
\affiliation{School of Physics and Astronomy, Yunnan University, Kunming 650091, P. R. China\\}
\affiliation{Key Laboratory of Astroparticle Physics of Yunnan Province, Yunnan University, Kunming 650091, P. R. China\\}

\author[0000-0001-7908-4996]{Benzhong Dai}
\affiliation{School of Physics and Astronomy, Yunnan University, Kunming 650091, P. R. China\\}
\affiliation{Key Laboratory of Astroparticle Physics of Yunnan Province, Yunnan University, Kunming 650091, P. R. China\\}

\email{bzhdai@ynu.edu.cn}

\begin{abstract}
Radiatively inefficient accretion flows (RIAFs) in low-luminosity active galactic nuclei (LLAGNs) are promising candidates for the PeV astrophysical diffuse neutrino background (DNB), but how the magnetically arrested disk (MAD) state affects their neutrino output and what fraction of LLAGNs are in this state remain unclear. We present a comprehensive study of diffuse neutrino emission from the inner accretion flows of LLAGNs in the MAD state, including all three flavors and both hadronuclear ($pp$) and photohadronic ($p\gamma$) production channels. Assuming that $\sim 5\%$ of the accretion power is transferred to nonthermal protons, current IceCube data allow at most $2.3\times10^{-3}$ of LLAGNs to have their inner accretion flows in a MAD state that can produce high-energy neutrinos. A lower fraction of the accretion power transferred to nonthermal protons would allow a proportionally larger fraction of such LLAGNs. Assuming that the comoving number density of low-excitation radio galaxies (LERGs) approximately matches that of LLAGNs in the MAD state, we also estimate a MAD fraction of $\simeq 1.6\times10^{-3}$. The corresponding diffuse intensity is compatible with current IceCube data and testable by IceCube-Gen2. For this MAD fraction, the all-flavor diffuse intensity peaks at $E_\nu\simeq1.5\,{\rm PeV}$ with $E_\nu^2\Phi_\nu\simeq2.6\times10^{-9}\,{\rm GeV\,cm^{-2}\,s^{-1}\,sr^{-1}}$. Current IceCube data are beginning to challenge several predictions from previous LLAGN RIAF models, and the MAD component can become the dominant contribution from LLAGN accretion flows beyond the cutoffs of most earlier RIAF spectra considered here.
\end{abstract}

\section{Introduction}

The IceCube Neutrino Observatory has established a diffuse astrophysical neutrino flux in the TeV--PeV range, commonly referred to as the diffuse neutrino background (DNB) \citep{2024PhRvD.110b2001A,2026PhRvL.136l1002A,2026PhRvD.113f2002A}. 
The origin of this diffuse flux remains uncertain, but extragalactic source populations are favored because the Galactic contribution is constrained to be subdominant \citep{2023Sci...380.1338I}.

Low-luminosity active galactic nuclei (LLAGNs) have low bolometric luminosities and low Eddington ratios, and their inner accretion flows are often interpreted as radiatively inefficient accretion flows (RIAFs) \citep{2008ARAA..46..475H,2014ARAA..52..529Y}. RIAFs in LLAGNs are promising environments for high-energy neutrino production. \citet{2015ApJ...806..159K} showed that LLAGNs can dominate the PeV diffuse neutrino intensity if about $1\%$ of the accretion power is carried by nonthermal protons. \citet{2021NatCo..12.5615K} proposed that LLAGNs can simultaneously explain the MeV gamma-ray background and the PeV neutrino background. \citet{2025ApJ...989..155K} showed with three-dimensional general relativistic magnetohydrodynamic (GRMHD) simulations that the global, inhomogeneous magnetohydrodynamic structure of RIAFs can affect the distribution of high-energy particles and the resulting neutrino emission, suggesting that the magnetic structure and state of the accretion flow should be taken into account when modeling high-energy neutrino production.

Accretion flows onto supermassive black holes are often classified as standard and normal evolution (SANE) or magnetically arrested disk (MAD), according to their magnetic state \citep{2003PASJ...55L..69N,2012MNRAS.426.3241N,Kimura2024ICRC}. Compared with the SANE state, the MAD state is characterized by stronger magnetic fields and by the accumulation of large-scale magnetic flux near the black hole \citep{2003PASJ...55L..69N,2011MNRAS.418L..79T,2020ApJ...905..178K}, which can significantly change the normalization and spectral shape of neutrino emission. \citet{2020ApJ...905..178K} showed that MADs associated with radio galaxies can produce PeV neutrinos and found that the accompanying diffuse gamma-ray emission contributes less than about $20\%$ of the extragalactic gamma-ray background in the $1$--$30\ \mathrm{GeV}$ range.

LLAGN RIAFs have long been considered leading candidates for the PeV DNB, but the fraction of LLAGN RIAFs in the MAD state and their resulting contribution to the DNB remain uncertain. Answering this question requires both the single-source neutrino spectra of MAD regions in LLAGNs and the MAD fraction within the LLAGN population, because the MAD contribution is set jointly by the neutrino yield of each MAD region and the number of LLAGNs hosting such regions. In this Letter, we present a comprehensive study of the diffuse neutrino spectrum from MAD regions in LLAGNs and compare our predictions with previous LLAGN RIAF models \citep{2015ApJ...806..159K,2021NatCo..12.5615K,2020ApJ...905..178K}. 

\section{Neutrino Production in MAD Regions}
\label{sec:mad_model}

We consider a neutrino-production region at $R=10R_{\rm g}$ in the inner accretion flow of an LLAGN, where $R_{\rm g}\equiv GM_{\rm BH}/c^2$. It is modeled as a hot RIAF in the MAD state, where accumulated magnetic flux strongly magnetizes the inner flow and affects its dissipation. The region contains nonthermal protons, thermal target protons, and thermal electrons, together with a photon field generated by the thermal electrons. Following previous RIAF and MAD models \citep{2015ApJ...806..159K,2019ApJ...874..168W,2020ApJ...905..178K}, we set the radial velocity and magnetic field as $V_r\simeq \alpha V_K/2$ and $B=\left(8\pi\rho C_s^2/\beta \right)^{1/2}$, where $\alpha=0.3$ and $\beta=0.1$ are the viscosity parameter and plasma beta, respectively. Here $V_K$, $C_s$, and $H$ denote the Keplerian velocity, sound speed, and scale height. We take $V_K=(GM_{\rm BH}/R)^{1/2}$, $C_s\simeq V_K/2$, $H/R\simeq C_s/V_K$, and the mass density then follows from mass conservation as $\rho\simeq \dot M/(4\pi R H V_r)$.

Assuming $n_e/n_p=1$ for the thermal plasma, the target gas density for hadronuclear interactions is $n_{\rm H}\equiv n_p\simeq \rho/m_p$. The dimensionless accretion rate is defined as $\dot m\equiv \dot M c^2/L_{\rm Edd}$, with $10^6\,M_\odot\leq M_{\rm BH}\leq 5 \times 10^{9}\,M_\odot$, $10^{-5}\leq\dot m\leq 3\times 10^{-2}$. All parameter sets satisfy $\tau_{\rm T}=n_{\rm H}\sigma_{\rm T}R<1$ and have subluminal Alfv\'en speeds $V_A=B/\sqrt{4\pi\rho}<c$.

Thermal electrons are assumed to follow a relativistic Maxwellian distribution. Following \citet{2020ApJ...905..178K}, the electron temperature is determined from the balance between electron heating and radiative cooling. The target photon field is calculated in the local comoving frame from thermal synchrotron emission, synchrotron self-absorption, bremsstrahlung, and inverse Compton scattering \citep{1997ApJ...477..585M,2015ApJ...806..159K,2020ApJ...905..178K}. The resulting differential photon number density is used for $p\gamma$ interactions and for estimating the $\gamma\gamma$ opacity. We find $\tau_{\gamma\gamma}\gg1$ at $E_\gamma\gtrsim0.1\,{\rm PeV}$ for all parameter sets considered, where $E_\gamma$ denotes the gamma-ray energy. MAD regions are therefore highly opaque to gamma rays in this energy range. The associated electromagnetic cascade in MADs has been treated in detail in \citet{2020ApJ...905..178K}.

The steady-state proton spectrum is obtained from
\begin{equation}
-\frac{d}{dE_p}\left[\dot E_p N_p(E_p)\right]
+\frac{N_p(E_p)}{t_{p,{\rm esc}}}=Q_p(E_p),
\label{eq:transport}
\end{equation}
where $N_p(E_p)$ is the differential spectrum of primary protons in the MAD region, $Q_p(E_p)$ is the injection rate, and $N_p(E_p)/V$ gives the differential number density, where $V$ is the volume of the MAD region. The proton energy-loss rate is defined as $\dot E_p=E_p/t_{p,{\rm cool}}$, with $t_{p,{\rm cool}}^{-1}=t_{pp}^{-1}+t_{p\gamma}^{-1}+t_{\rm BH}^{-1}+t_{p,{\rm syn}}^{-1}$ for inelastic $pp$ collisions, photomeson production, Bethe--Heitler pair production, and proton synchrotron cooling \citep{2019MNRAS.485..163K,2020ApJ...905..178K}.

Protons can leave the MAD region either by radial advection toward the black hole or by diffusion out of the region \citep{2019MNRAS.485..163K,2020ApJ...905..178K}, $t_{p,{\rm esc}}^{-1}=t_{\rm fall}^{-1}+t_{\rm diff}^{-1}$, where $t_{\rm fall}\simeq \frac{R}{V_r}$ and $t_{\rm diff}\simeq \frac{R^2}{D_R}$. For diffusive escape, we adopt $D_R\simeq cE_p/(3eB)$, corresponding to Bohm diffusion.

The proton injection rate is assumed to be $Q_p(E_p)=Q_{0,p}\left(E_p/E_{p,0}\right)^{-s}
\exp\left(-E_p/E_{p,{\rm cut}}\right)$, with $s=1.3$ following \citet{2020ApJ...905..178K}. The nonthermal proton injection is treated phenomenologically and may arise from magnetic reconnection or turbulence in the MAD region. The normalization, $Q_{0,p}$, is determined by
\begin{equation}
L_p=\epsilon_{\rm NT}\epsilon_{\rm dis}\dot M c^2 =\int_{m_pc^2}^{\infty} dE_p\,E_p Q_p(E_p),
\label{eq:proton_power}
\end{equation}
where $\epsilon_{\rm dis}=0.15$ is the energy fraction of the accretion rest-mass power dissipated in the MAD region, and $\epsilon_{\rm NT}=0.33$ is the energy fraction of the dissipated power injected into nonthermal protons \citep{2020ApJ...905..178K}. The cutoff energy is set by $E_{p,{\rm cut}}=\min[E_{p,{\rm loss}},E_{p,{\rm Hillas}}]$, where the Hillas energy is estimated as $E_{p,{\rm Hillas}}\simeq eBRV_A/(\eta_{\rm acc}c)$, and $E_{p,{\rm loss}}$ is obtained from $t_{p,{\rm acc}}(E_p)=t_{p,{\rm loss}}(E_p)$, with the total loss timescale $t_{p,{\rm loss}}^{-1}=t_{p,{\rm cool}}^{-1}+t_{p,{\rm esc}}^{-1}$. We use the acceleration time $t_{p,{\rm acc}}\simeq \eta_{\rm acc}E_p(eBc)^{-1}(c/V_A)^2$ with $\eta_{\rm acc}=5$.  
This conservative choice avoids maximizing $E_{p,{\rm cut}}$ and reduces the sensitivity of the PeV neutrino component to the highest-energy proton tail.

We calculate neutrino production from the $pp$-pion ($pp\to\pi^\pm+X$), $pp$-kaon ($pp\to K^\pm+X$), $pp$-charm ($pp\to c\bar{c}+X$), $p\gamma$-pion ($p\gamma\to\pi^\pm+X$), and $p\gamma$-kaon ($p\gamma\to K^\pm+X$) channels. The secondary $\pi^\pm$, $K^\pm$, and $\mu^\pm$ spectra are obtained from transport equations analogous to Eq.~\ref{eq:transport}, including synchrotron cooling, escape, and decay. For $\pi^\pm$ and $K^\pm$, we also include inelastic meson--proton scattering with $t_{hp}^{-1}=\kappa_{hp}n_{\rm H}c\sigma_{hp}$, where $\kappa_{hp}\simeq0.8$ and $\sigma_{hp}\sim10^{-25}\,{\rm cm}^2$ \citep{2009PhRvD..79e3006E,2020PhRvD.102j3001C,2025arXiv251221805W,2024MNRAS.528L..88Z}. The two-body decays of charged pions and kaons are included, and neutrinos from the subsequent three-body muon decay are calculated using distributions that include muon polarization \citep{2007PhRvD..75l3005L,2009PhRvD..79e3006E,2015JCAP...06..034B,2020PhRvD.102j3001C}.

The $p\gamma$ channels are treated with SOPHIA \citep{2000CoPhC.124..290M,2010ApJ...721..630H}, and the resulting charged-pion and charged-kaon final-state distributions are transformed to the local comoving frame. For the $pp$ channels, the inclusive differential production cross sections for $\pi^\pm$, $K^\pm$, and charmed hadrons are constructed from SIBYLL~2.3e through the Chromo interface \citep{2019PhRvD.100j3018F,2020PhRvD.102f3002R,2026CoPhC.32110031F}. We include $D^0$, $D^\pm$, $D_s^\pm$, and $\Lambda_c$, together with their antiparticles. Since charmed hadrons decay much faster than the relevant cooling and escape times, their cooling and escape are neglected before decay. We vary the charm-hadron yields between $1/3$ and $3$ of the fiducial value to estimate the charm-production uncertainty \citep{2020PhRvD.102j3001C}. For each neutrino flavor, the spectra from these decay chains are summed over all production channels to obtain the single-source differential neutrino production rate $Q_{\nu_\alpha}(E_\nu';\dot m,M_{\rm BH})$, where $\alpha=e,\mu,\tau$.

\section{Diffuse Neutrino Intensity from the LLAGN Population}
\label{sec:diffuse_intensity}

To compute the diffuse neutrino intensity, we interpolate the single-source differential neutrino production rate $Q_{\nu_\alpha}(E_\nu';\dot m,M_{\rm BH})$ as a function of the bolometric luminosity $L_{\rm bol}(\dot m,M_{\rm BH})$, which is obtained by integrating the thermal photon spectrum over all frequencies for each $(\dot m,M_{\rm BH})$. At each $L_{\rm bol}$, we average over black-hole mass using the supermassive black-hole mass function (BHMF) of \citet{2004MNRAS.354.1020S} as a normalized relative weight, obtaining the mass-weighted differential neutrino production rate $\bar Q_{\nu_\alpha}(E_\nu',L_{\rm bol})$. We assume that, at fixed $L_{\rm bol}$, the relative distribution of black-hole masses follows the adopted BHMF. The BHMF sets the relative contributions from different black-hole masses, whereas the LLAGN luminosity function sets the source density. We then integrate $\bar Q_{\nu_\alpha}[(1+z)E_\nu,L_{\rm bol}]$, with $E_\nu'=(1+z)E_\nu$, over the LLAGN luminosity function, and the all-flavor diffuse intensity is
\begin{equation}
\begin{aligned}
\Phi_\nu(E_\nu)=&\frac{c}{4\pi}
\int_0^{z_{\rm max}}\frac{dz}{H(z)} \\
&\times\int d\log_{10}L_{\rm bol}\,
f_{\rm MAD}\,\phi_{\rm bol}(L_{\rm bol}) \\
&\times \sum_{\alpha=e,\mu,\tau}\bar Q_{\nu_\alpha}[(1+z)E_\nu,L_{\rm bol}].
\end{aligned}
\label{eq:diffuse_intensity}
\end{equation}

Here $\phi_{\rm bol}(L_{\rm bol})\equiv dN/(dV_c\,d\log_{10}L_{\rm bol})$ is the LLAGN bolometric luminosity function per logarithmic luminosity interval, where $N$ is the number of sources and $V_c$ is the comoving volume. We obtain this function by converting the H$\alpha$ luminosity function of \citet{2015ApJ...806..159K} to logarithmic form and evaluating the converted function at $L_{\rm H\alpha}=L_{\rm bol}/80$. The adopted H$\alpha$ luminosity interval $3\times10^{36}$--$3\times10^{41}\,{\rm erg\,s^{-1}}$ corresponds to $L_{\rm bol}=2.4\times10^{38}$--$2.4\times10^{43}\,{\rm erg\,s^{-1}}$. We neglect redshift evolution of the luminosity function and set $z_{\rm max}=5$. We use $H(z)=H_0[\Omega_m(1+z)^3+\Omega_\Lambda]^{1/2}$ with $H_0=67.7\,{\rm km\,s^{-1}\,Mpc^{-1}}$, $\Omega_m=0.31$, and $\Omega_\Lambda=0.69$ \citep{2020A&A...641A...6P}. The flavor composition at Earth is obtained from that at the source using long-baseline averaged vacuum oscillations 
\citep{2024JHEP...12..216E}.

The key parameter is $f_{\rm MAD}$, defined here as the fraction of LLAGNs whose inner accretion flows are in the modeled MAD state. We assume that $f_{\rm MAD}$ is independent of luminosity, black hole mass, and redshift, so that it linearly scales the predicted diffuse neutrino intensity. Low-excitation radio galaxies (LERGs) are radio-mode AGNs and are commonly associated with RIAFs \citep{2012MNRAS.421.1569B,2014ARA&A..52..589H}. They have also been used to represent the MAD source population in previous population calculations \citep{2020ApJ...905..178K}. Following \citet{2014ARA&A..52..589H,2020ApJ...905..178K}, we integrate the $1.4\,{\rm GHz}$ LERG radio luminosity function over the monochromatic radio power range $\nu P_\nu=10^{38.15}$--$10^{42.65}\,{\rm erg\,s^{-1}}$, and obtain the comoving number density $n_{\rm LERG}\simeq7.5\times10^{-5}\,{\rm Mpc^{-3}}$. Integrating the H$\alpha$ LLAGN luminosity function adopted from \citet{2015ApJ...806..159K} gives $n_{\rm LLAGN}\simeq4.7\times10^{-2}\,{\rm Mpc^{-3}}$. Assuming that the comoving number density of LERGs approximately matches that of LLAGNs in the MAD state, we estimate the MAD fraction from the ratio of the LERG and LLAGN number densities derived from their respective luminosity functions and denote this estimate as $f_{\rm MAD}^{\rm LF}$, giving $f_{\rm MAD}^{\rm LF}\simeq n_{\rm LERG}/n_{\rm LLAGN}\simeq1.6\times10^{-3}$. This estimate is independent of the neutrino model, and the corresponding diffuse intensity can be compared with IceCube observations.

\begin{figure}[t] 
\centering
\includegraphics[width=\linewidth]{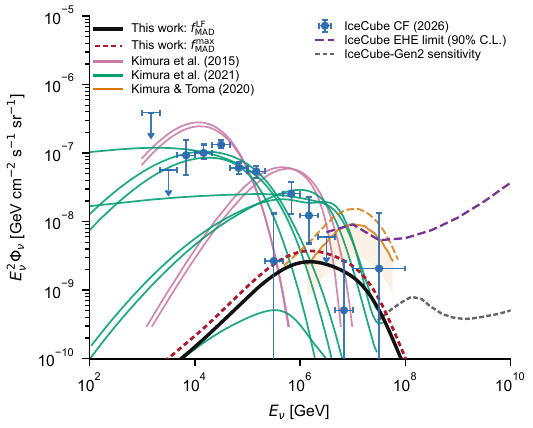}
\caption{All-flavor diffuse neutrino intensities from this work and previous LLAGN accretion-flow models. The black solid and red dotted curves show the diffuse neutrino intensities predicted in our work for $f_{\rm MAD}=f_{\rm MAD}^{\rm LF}$ and $f_{\rm MAD}=f_{\rm MAD}^{\rm max}$, respectively. Thin solid curves in magenta and green show the LLAGN RIAF model predictions of \citet{2015ApJ...806..159K} and \citet{2021NatCo..12.5615K}, respectively. Orange solid and dashed curves show the MAD model predictions of \citet{2020ApJ...905..178K}. The blue markers show the all-flavor IceCube Combined Fit (CF) segmented spectrum, with vertical error bars and downward arrows denoting the reported 68\% intervals and upper limits, respectively \citep{2026PhRvD.113f2002A}. The purple dashed curve shows the IceCube extremely-high-energy (EHE) differential upper limit at the 90\% confidence level \citep{2025PhRvL.135c1001A}. The gray dashed curve shows the projected IceCube-Gen2 sensitivity \citep{2021JPhG...48f0501A,2022JHEAp..36...55A}.}
\label{fig:population_summarya}
\end{figure}

Figure~\ref{fig:population_summarya} shows previous LLAGN RIAF models, this work, and current IceCube data. \citet{2015ApJ...806..159K} vary the black-hole mass, electron-heating fraction, turbulence parameter, and proton injection efficiency. \citet{2021NatCo..12.5615K} use different H$\alpha$ luminosity functions and proton-acceleration prescriptions. \citet{2020ApJ...905..178K} show solid and dashed curves for the jet production efficiency $\epsilon_j=1$ and $0.6$, respectively. The orange shaded region indicates the uncertainty of a factor of 3 by radio luminosity functions. Differences in the adopted source population and black-hole mass distribution, together with differences in the individual-source neutrino emission model, affect both the peak energy and the diffuse neutrino intensity. In contrast, $f_{\rm MAD}$ only rescales the diffuse intensity and does not affect the spectral shape or peak energy. Both our model and that of \citet{2020ApJ...905..178K} associate MADs with LERGs, but use the LERG population differently. The earlier model integrates a LERG radio luminosity function, converts radio luminosity to accretion rate through a radio--jet-power relation, and adopts $M_{\rm BH}=10^9\,M_\odot$. In our work, the LERG number density sets $f_{\rm MAD}^{\rm LF}$, which is then applied to the LLAGN H$\alpha$ luminosity function while the source spectra are averaged over the BHMF. This allows IceCube data to assess how common the MAD state can be among LLAGN RIAFs.

Comparison with current IceCube data gives an allowed maximum value $f_{\rm MAD}^{\max}\simeq 2.3\times10^{-3}\left(\frac{0.33}{\epsilon_{\rm NT}}\right)\left(\frac{0.15}{\epsilon_{\rm dis}}\right)$. For $f_{\rm MAD}=f_{\rm MAD}^{\rm LF}=1.6\times 10^{-3}$, the predicted diffuse intensity is compatible with the IceCube CF segmented spectrum, remains below the IceCube EHE differential upper limit, and is within reach of IceCube-Gen2 near $E_\nu\sim30\,{\rm PeV}$. At this normalization, the all-flavor intensity peaks at $E_\nu\simeq1.5\,{\rm PeV}$ with $E_\nu^2\Phi_\nu\simeq2.6\times10^{-9}\,{\rm GeV\,cm^{-2}\,s^{-1}\,sr^{-1}}$.

Updated IceCube data are beginning to challenge several previously viable models, including the higher-intensity cases of the earlier MAD model. The highest cutoff energies among the previous SANE RIAF spectra considered here are $E_\nu\sim10$--$30\,{\rm PeV}$. The exact spectral contrast also depends on the particle-acceleration and population assumptions, but the comparison shows that the MAD component extends the LLAGN accretion-flow neutrino spectrum to higher energies and can become dominant beyond the earlier cutoffs.

\begin{figure}[t]
\centering
\includegraphics[width=\linewidth]{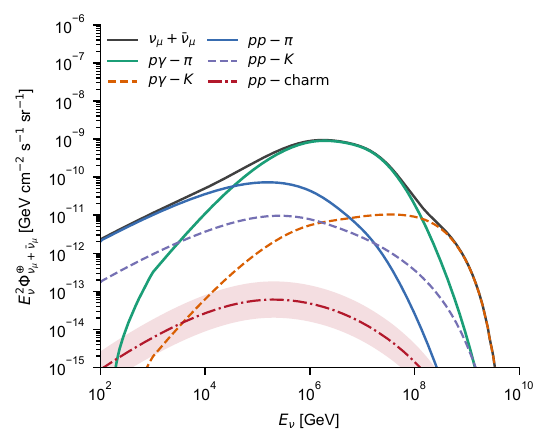}
\caption{Earth-frame $\nu_\mu+\bar\nu_\mu$ intensity separated by production channel for $f_{\rm MAD}=f_{\rm MAD}^{\rm LF}$; the shaded band varies the charm-hadron yields by factors of $1/3$ and $3$.}
\label{fig:population_summaryb}
\end{figure}

The peak in Figure~\ref{fig:population_summaryb} is dominated by $p\gamma$-pion production, which accounts for about $95\%$ of the total Earth-frame $\nu_\mu+\bar\nu_\mu$ intensity. The $pp$-pion channel is more visible at lower energies, while $p\gamma$-kaon production becomes more important in the high-energy tail. The $pp$-charm component remains negligible even at the upper edge of the shaded band. Around this peak, flavor mixing converts the source-frame ratio $(\Phi_{\nu_e}:\Phi_{\nu_\mu}:\Phi_{\nu_\tau})_{\rm src}\simeq(0.240:0.760:0)$ to the Earth-frame ratio $(\Phi_{\nu_e}:\Phi_{\nu_\mu}:\Phi_{\nu_\tau})_{\oplus}\simeq (0.296:0.361:0.344)$, with neutrinos and antineutrinos summed. These relative channel contributions and flavor ratios remain unchanged as $f_{\rm MAD}$ varies, since $f_{\rm MAD}$ only rescales the overall diffuse intensity.


\section{Summary and Discussion}
\label{sec:discussion_conclusions}

We examined neutrino production in MAD regions of LLAGNs, resolving the neutrino spectra by production channel and flavor, and estimated the resulting diffuse neutrino spectrum. We use $f_{\rm MAD}^{\rm LF}\simeq1.6\times10^{-3}$ to set the normalization of the predicted diffuse intensity. The all-flavor intensity peaks at $E_\nu\simeq 1.5\,{\rm PeV}$ with $E_\nu^2\Phi_\nu\simeq2.6\times10^{-9}\,{\rm GeV\,cm^{-2}\,s^{-1}\,sr^{-1}}$. The corresponding diffuse intensity is compatible with the updated CF segmented spectrum, remains below the EHE differential upper limit, and can be tested by IceCube-Gen2.

The peak is dominated by $p\gamma$-pion production on optical/UV target photons generated through inverse Compton scattering by thermal electrons, because the target photon density is relatively high. Below $E_\nu\simeq0.1\,{\rm PeV}$, the corresponding target photons shift to higher energies, where the photon density is lower, suppressing the $p\gamma$ efficiency. The $pp$-pion component from the dense gas therefore dominates in this energy range, while the $pp$-charm contribution remains negligible. In the high-energy tail, pion and muon cooling suppress the pion--muon decay chain, whereas kaons are less affected because of their larger mass and shorter lifetime $\tau$, making the $p\gamma$-kaon component the main contribution there. Near the peak, the flavor composition after propagation is $(\Phi_{\nu_e}:\Phi_{\nu_\mu}:\Phi_{\nu_\tau})_\oplus\simeq(0.296:0.361:0.344)$ after summing neutrinos and antineutrinos, with the $\nu_\mu+\bar\nu_\mu$ fraction slightly above one third. This excess mainly reflects partial muon damping, which modifies the source flavor composition before oscillations.

The comparison with current IceCube data gives an allowed maximum fraction of LLAGNs hosting MAD states that can produce high-energy neutrino emission,
\begin{equation}
f_{\rm MAD}^{\rm max}\simeq 2.3\times10^{-3}\left(\frac{0.0495}{\epsilon_{\rm NT}\epsilon_{\rm dis}}\right).
\end{equation}
Within our model, the IceCube comparison constrains the product $f_{\rm MAD}\epsilon_{\rm NT}\epsilon_{\rm dis}$, not $f_{\rm MAD}$ alone. Here, $\epsilon_{\rm NT}\epsilon_{\rm dis}$ is the proton-power fraction, defined as the fraction of the accretion power supplied to nonthermal protons in each source. We adopt $\epsilon_{\rm NT}\epsilon_{\rm dis}=0.0495\simeq5\%$, as in previous MAD models \citep{2020ApJ...905..178K,2022ApJ...935..159K,2024ApJ...977...22K}. A lower proton-power fraction would allow a proportionally larger $f_{\rm MAD}^{\max}$.

Equivalently, $f_{\rm MAD}^{\rm max}$ can be viewed as the maximum allowed probability for an LLAGN RIAF to be in such a MAD state. A lower proton-power fraction would allow a proportionally larger probability. GRMHD simulations find that approximately $15\%-40\%$ of the magnetic flux supplied at large radii can reach the black hole \citep{2023ApJ...944..182D}. This suggests that, once sufficient large-scale poloidal magnetic flux is available, inward transport through the RIAF need not be the main bottleneck. Net poloidal magnetic flux can then accumulate near the black hole until the associated magnetic pressure becomes strong enough to impede the inflow and drive the system into the MAD state \citep{2011MNRAS.418L..79T}. Sustained accumulation requires the supplied poloidal flux to remain predominantly of one polarity over the time needed to build up the magnetic flux threading the black hole. If the probability for an LLAGN to be in such a MAD state is indeed only a few per thousand, this may indicate that the conditions required for an LLAGN to acquire sufficient coherent net poloidal magnetic flux and retain it long enough to sustain a MAD are rarely satisfied.


Updated IceCube data are beginning to challenge several previously viable models. Our work connects the high-energy DNB to the magnetic state of LLAGN accretion flows. The MAD component remains compatible with current IceCube data, can dominate the LLAGN accretion-flow neutrino spectrum beyond the cutoffs of most earlier RIAF spectra, and is within reach of IceCube-Gen2.

\begin{acknowledgments}
We thank Dr. Jimmy Hammer for helpful discussions on secondary-particle production and decay. We thank Jiawen Li of Yunnan University for helpful discussions on the fraction of LLAGNs in the MAD state. The numerical computations were conducted on the Yunnan University Astronomy Supercomputer. This work was partially supported by the National Science Foundation of China (12263007 and 12233006), and by the High-level Talent Support Program of Yunnan Province.
\end{acknowledgments}

\bibliographystyle{aasjournal}
\bibliography{references}

\end{document}